\documentclass[a4paper,11pt]{article}
\pdfoutput=1 
\usepackage{jheppub}
\usepackage[T1]{fontenc}
\usepackage{amsmath}
\usepackage{mathtools}
\usepackage{subcaption}
\newcommand{\Z}{\mathbb{Z}}
\newcommand{\SU}{\mathrm{SU}}
\newcommand{\U}{\mathrm{U}}

\newcommand{\re}{{\mathrm{Re}}}
\newcommand{\dd}{{\rm{d}}}
\newcommand{\SW}{S_{\mbox{\tiny{W}}}}

\newcommand{\redchisq}{\chi^2_{\tiny\mbox{red}}}

\title{\boldmath Width of the flux tube in compact $\U(1)$ gauge theory in three dimensions}

\author{Michele Caselle,}
\author{Marco Panero}
\author{and Davide Vadacchino}

\affiliation{Department of Physics, University of Turin \& INFN, Turin \\
Via Pietro Giuria 1, I-10125 Turin, Italy}

\emailAdd{caselle@to.infn.it}
\emailAdd{marco.panero@unito.it}
\emailAdd{vadacchi@to.infn.it}

\abstract{We study the squared width and the profile of flux tubes in compact $\U(1)$ lattice gauge theory in three spacetime dimensions. The results obtained from numerical calculations in the dual formulation of this confining theory are compared with predictions from an effective bosonic-string model and from the dual-superconductor model: it is found that the former fails at describing the quantitative features of the flux tube, while the latter is in good agreement with Monte Carlo data. The analytical interpretation of these results (in the light of the semi-classical analysis by Polyakov) is pointed out, and a comparison with non-Abelian gauge theories in four spacetime dimensions is discussed.}


\arxivnumber{1601.07455}

\begin{document}

\maketitle
\flushbottom

\section{Introduction}
\label{sec:introduction}

One of the most interesting features of confining gauge theories is that flux lines are concentrated into thin, fluctuating tubes: this is clearly seen in numerical studies on the lattice~\cite{Bali:1994de, Bali:2000gf, Takahashi:2000te, Bissey:2006bz}. During the past few years, significant computational and analytical efforts have been devoted to a more precise characterization of the flux tube. The main reason is that its behavior is deeply related to the underlying confining mechanism and it is hoped that one could shed some light on this issue by studying the physics of the flux tube.

In particular, there are two seemingly conflicting descriptions for the flux tube, respectively based on an effective string theory (EST) approach and on the dual-superconductor model. The former is obtained within the framework of the effective description of confinement~\cite{Luscher:1980fr} in terms of a Nambu-Got{\={o}} string~\cite{Nambu:1974zg, Goto:1971ce} (see ref.~\cite{Aharony:2013ipa} for a discussion of recent theoretical developments, and ref.~\cite{Lucini:2012gg} for a review of numerical results): for a theory defined in $d$ spacetime dimensions, it predicts that at low temperatures the squared width of the flux tube $w^2$ increases logarithmically with the distance $R$ between the color sources~\cite{Luscher:1980iy}:
\begin{equation}
w^2 = \frac{d-2}{2\pi\sigma} \ln \left( \frac{R}{R_0} \right),
\label{f1}
\end{equation}
where $d-2$ is the number of transverse dimensions along which the flux tube vibrates, $\sigma$ is the string tension, and $R_0$ is a parameter with the dimensions of a length. This prediction has been confirmed by numerical simulations in different confining lattice gauge theories (LGTs), both with Abelian~\cite{Caselle:1995fh, Zach:1997yz, Koma:2003gi, Panero:2005iu, Giudice:2006hw, Rajantie:2012zn, Amado:2013rja} and non-Abelian local symmetry~\cite{Pennanen:1997qm, Chernodub:2007wi, Bakry:2010zt, Cardoso:2013lla}.

In the effective string framework, eq.~(\ref{f1}) corresponds to the first (or ``Gau{\ss}ian'') order of approximation in an expansion around the long-string limit for a ``mesonic'' flux tube (see also refs.~\cite{Alexandrou:2001ip, deForcrand:2005vv, Pfeuffer:2008mz, Bakry:2015csa, Bakry:2014gea}, for analogous studies in a ``baryonic'' setup). This analytical computation can be extended to the next-to-leading order~\cite{Gliozzi:2010zv, Gliozzi:2010zt} and remarkable agreement with Monte~Carlo results has been found also at this level~\cite{Gliozzi:2010zv, Gliozzi:2010zt, Caselle:2010zs, Gliozzi:2010jh}.

At finite temperature $T$ (but still in the confining regime), the same effective theory predicts a linear increase of the squared width, with a proportionality constant that diverges as the deconfinement temperature is approached from below~\cite{Allais:2008bk, Caselle:2010zs}. Also these predictions were recently confirmed by numerical simulations~\cite{Caselle:2010zs, Gliozzi:2010jh}. 

Until now, the predictions of the effective string model for the width of confining flux tubes have always been confirmed by numerical lattice results. In this work, however, we will show that the $\U(1)$ lattice gauge theory in three dimensions provides a significant exception to this rule of thumb. More precisely, we will demonstrate that the standard string picture does not provide a good low-energy description for this model, and that sizeable deviations from eq.~(\ref{f1}) can be observed in lattice simulations of this theory.\footnote{Although the flux-tube width in this lattice theory has been studied numerically since the 1980's~\cite{Sterling:1983fs}, the computing-power limitations of the time did not allow one to reach the level of precision required for a detailed quantitative comparison with the theoretical model.} Na\"{\i}vely, one could suspect that this mismatch is just due to the fact that the effective string description of the confining flux tube holds only for interquark distances larger than $1/\sqrt{\sigma}$. However, we will show that this is not the case, and that the disagreement has a more subtle explanation. 

The fact that compact $\U(1)$ lattice gauge theory in three dimensions (3D) exhibits a behavior different from other confining theories should not come as a surprise. This model, which has been investigated analytically since forty years ago~\cite{Polyakov:1976fu, Banks:1977cc, Drell:1978hr, Gopfert:1981er}, has the interesting property that two characteristic non-perturbative, dimension-$1$ quantities, namely $m_0$ (the mass of the lightest ``glueball'') and $\sqrt{\sigma}$ (the square root of the asymptotic slope of the potential $V(R)$ associated with two static charges at a distance $R$ from each other), scale differently with the coupling $e$---or, equivalently, with the lattice spacing $a$---of the lattice theory.\footnote{Note that this is a property of the lattice theory at finite spacing, it does \emph{not} mean that the continuum theory is characterized by two independent scales.} This can be contrasted with the behavior of $\SU(N)$ lattice gauge theories, in which the dependence of the $m_0/\sqrt{\sigma}$ ratio on the  value of the coupling is a very mild one (and simply due to discretization artifacts that vanish in the continuum limit). As discussed in our previous publication~\cite{Caselle:2014eka}, this peculiar property of compact $\U(1)$ lattice gauge theory in 3D has a direct counterpart at the level of the effective string model describing its long-distance physics, whose action includes an extrinsic-curvature contribution~\cite{Orland:1994qt}, besides the usual Nambu-Got\={o} term. The extrinsic-curvature term means that the flux tube vibrates like a ``rigid'' string~\cite{Caselle:2014nta}; the presence of contributions of this type (which are compatible with Lorentz symmetry~\cite{Aharony:2011gb}) also in the effective string action for $\SU(N)$ gauge theories is not ruled out \emph{a priori}; however, numerically, it seems to be hard to detect, within the precision of state-of-the-art lattice studies~\cite{Brandt:2013eua}. The fact, that in compact lattice QED in 3D the effect of extrinsic-curvature terms can have a sizeable impact on the infrared dynamics, makes this theory particularly attractive for numerical studies of the confining string model~\cite{Vadacchino:2013baa}. This motivated the analysis of the confining potential in this theory, that we reported in refs.~\cite{Caselle:2014eka, Vadacchino:2014ota}. In the present work, we extend this investigation to the width and the profile of the flux tube.

Coming back to eq.~(\ref{f1}), our lack of knowledge about the flux-tube dynamics at short distances is encoded in the $R_0$ parameter, which cannot be predicted from the string model alone, and sets a lower bound on the distances at which the effective string description is expected to hold. In the effective string model, the flux tube is idealized as a one-dimensional object, whose typical width is purely due to quantum fluctuations. However, a more realistic picture of the flux tube may include a core of finite radius, whose size defines an ``intrinsic width'' of the flux tube.

The importance of this intrinsic width can be appreciated by probing the shape of the flux tube (by monitoring some local observable like, e.g., the field strength) along the transverse directions: moving from the axis joining the color sources to larger and larger transverse distances, the effective string picture predicts a Gau{\ss}ian profile, but the results of Monte~Carlo simulations reveal significant deviations from this idealized shape. Intriguingly, however, the second moment of the distribution is fully consistent with eq.~(\ref{f1}), if $R>R_0$.

A different way to model the flux tube in confining gauge theories is based on the dual-superconductor picture---an idea that can be traced back to seminal works by Nambu~\cite{Nambu:1974zg}, by 't~Hooft~\cite{'tHooft:1979uj} and by Mandelstam~\cite{Mandelstam:1974pi} (see ref.~\cite{Baker:1991bc} for a review and refs.~\cite{Cea:1992sd, Cea:1992vx, Cea:1993pi, Cea:1994ed, Cea:1994aj, Cea:1995zt, Cea:1995ga, Cardoso:2013lla, Amado:2013rja, Cardaci:2010tb, Cea:2012qw, Cea:2014uja} for comparisons with Monte~Carlo simulations). In this framework, quark confinement is associated with condensation of chromomagnetic monopoles, in analogy with Cooper-pair condensation in superconductors, so that the flux tube is expected to behave as a (dual) Abrikosov vortex. In contrast to the effective string model, the dual-superconductor picture predicts that the flux density decreases exponentially with the distance from the interquark axis (the associated  characteristic length scale being the analogue of the ``penetration length'' of Abrikosov vortices in a superconductor) and, more importantly, that it does not depend on the interquark distance $R$. These features can be easily associated with the rigid core mentioned above, with the penetration length playing the r\^ole of the intrinsic width. 

In the past few years there have been some attempts to combine the two pictures into one framework, merging an Abrikosov-like short-distance description of the flux-tube core with a long-distance effective string description of its quantum fluctuations~\cite{Cardoso:2013lla, Amado:2013rja}. This requires constructing an expression for the flux-tube shape, which should interpolate between the Gau{\ss}ian behavior predicted by the effective string and the exponential decrease predicted by the dual-superconductivity model. This interpolation can be carried out without arbitrary assumptions, when the physics allows one to use some additional theoretical tools. For example, it is known that in the presence of a continuous thermal deconfinement transition, the long-distance properties of a gauge theory should be described by a lower-dimensional spin model~\cite{Svetitsky:1982gs}: taking advantage of this, in refs.~\cite{Caselle:2006wr, Caselle:2012rp} the analytical study of the flux-tube width in LGTs in three dimensions was mapped to a problem in perturbed conformal field theory in two dimensions. Another theoretical tool, that can be used to study the flux-tube profile (at least in the strong-coupling limit) is the holographic correspondence~\cite{Maldacena:1997re, Gubser:1998bc, Witten:1998qj}: this approach was followed in refs.~\cite{Vyas:2010wg, Giataganas:2015yaa}.

Some recent works tried to work out a model interpolating between the thin-string and the dual-superconductor-tube pictures also at zero (or low) temperature: they are inspired by the dynamics of Abrikosov vortices in superconductivity~\cite{Cea:2014uja}, or on the convolution of the classical intrinsic-width string behavior with a Gau{\ss}ian distribution, which should model the quantum oscillations of the flux tube far from its core~\cite{Cardoso:2013lla}. On the other hand, an interesting alternative proposal to describe the flux-tube shape at low temperature $T$ was inspired by the physics of rough interfaces: in particular, in a series of articles on this subject, M\"unster and collaborators derived the interface profile at one loop, in the framework of the renormalized $\phi^4$ model in three dimensions~\cite{Munster:1990yg, Klessinger:1992qq, Hoppe:1997ey, Muller:2004vv, Kopf:2008hr}.

As we will discuss in the following sections, the results of the high-precision Monte~Carlo study that we carried out reveal a wealth of interesting information about the confining flux tube in compact $\U(1)$ lattice gauge theory in 3D---in particular about its shape (as a function of the distance from the axis of the static sources) and width. Although numerical studies of this type are comparatively challenging, in this work we could reach high precision, thanks to an exact duality transformation of the Kramers-Wannier type~\cite{Kramers:1941kn, Wegner:1984qt, Cobanera:2011wn, Wegner:2014ixa}, which maps the original gauge theory to a spin model with integer-valued variables and nearest-neighbor interactions only. The main finding of this work is that a bosonic Nambu-Got\={o} string does not provide a quantitatively fully satisfactory model of flux tubes in this lattice theory (although it captures their main features at the qualitative and semi-quantitative level). Looking at the flux-tube profile at large transverse separations from the sources' axis, we also find that some of its features can be described well in terms of a dual Abrikosov vortex. We remark that, although the theory that we are considering has a finite ultraviolet cutoff (and some of its features---including, in particular, the existence of two independent dimensionful non-perturbative scales $m_0$ and $\sigma$---depend crucially on the finiteness of $a$), these results provide novel insight into the implications of confinement for the low-energy properties of a theory, and can serve as a guide to a better understanding of confinement in non-Abelian gauge theories in four spacetime dimensions, too.

The structure of this article is the following: in section~\ref{sec:simulation_setup}, we introduce the basic definitions of the $\U(1)$ model in three dimensions and of its lattice regularization, some of its interesting physical properties, and the setup of our Monte~Carlo study of the flux tube in this theory. In section~\ref{sec:results}, we present and analyze the numerical results of our lattice simulations. Finally, in section~\ref{sec:discussion_and_concluding_remarks} we summarize the implications of our findings, and point out some concluding remarks. 

\section{Simulation setup}
\label{sec:simulation_setup}

The model we are interested in is compact $\U(1)$ gauge theory in three spacetime dimensions, regularized on a Euclidean, isotropic, cubic lattice $\Lambda$ of spacing $a$. The action of the lattice theory is taken to be the Wilson action~\cite{Wilson:1974sk}
\begin{equation}
\label{Wilson_action}
\SW = \frac{1}{a e^2} \sum_{x \in \Lambda} \sum_{1 \le \mu < \nu \le 3} \left[1 - \re \, U_{\mu\nu}(x)\right],
\end{equation}
where $e$ denotes the coupling and $U_{\mu\nu}(x)$ is a plaquette:
\begin{equation}
\label{plaquette}
U_{\mu\nu}(x) = U_\mu(x) U_\nu(x+a\hat\mu) U^\star_\mu(x+a\hat\nu) U^\star_\nu(x).
\end{equation}
$U_\mu(x)$ denotes the parallel transporter defined on the oriented bond joining the nearest-neighbor lattice sites $x$ and $x + a \hat{\mu}$: it can be related to the continuum gauge field $A$ as $U_\mu(x) = \exp \left[ i a A_\mu \left( x + a \hat{\mu}/2 \right) \right]$. Thus, the partition function of the theory reads
\begin{equation}
\label{lattice_partition_function}
Z = \int \prod_{x \in \Lambda} \prod_{\mu=1}^3 {\dd}U_\mu(x) \, e^{-\SW},
\end{equation}
where ${\dd}U_\mu(x)$ denotes the Haar measure for $U_\mu(x)$. As shown in refs.~\cite{Polyakov:1976fu, Gopfert:1981er}, this theory admits a semi-classical solution, from which one can derive that the model is confining at any value of $\beta=1/(ae^2)$, and that for $ \beta\gg 1$ it can be described as a model of free, massive scalars.

As already pointed out in ref.~\cite{Caselle:2014eka}, a remarkable feature of this lattice theory is that the ratio between the mass of the lightest ``glueball'' ($m_0$) and the square root of the string tension has a strong (exponential) dependence on the coupling, so that, by varying $\beta$, the $m_0/\sqrt{\sigma}$ ratio of the lattice theory can be tuned to any arbitrary value.

Being an Abelian LGT in three dimensions, this theory can be reformulated as a 3D spin model, by means of an exact duality transformation~\cite{Kramers:1941kn, Wegner:1984qt, Cobanera:2011wn, Wegner:2014ixa}.\footnote{Interestingly, this type of duality transformations has recently been applied also in the numerical study of systems affected by a computational sign problem~\cite{Gattringer:2014nxa}.} Under this mapping, correlators of Polyakov loops in the original formulation of the theory can be rewritten in terms of a modified partition function of the dual spin system, involving a set of frustrations on the links dual to a surface bordered by the Polyakov loops. Using this duality transformation, reliable numerical evidence of the logarithmic growth of the squared width of the flux tube in the 3D $\Z_2$ gauge model was already obtained more than twenty years ago~\cite{Caselle:1995fh}. This approach can be easily applied in the $\U(1)$ model and combined with the factorization underlying the ``snake algorithm''~\cite{deForcrand:2000fi}, as discussed in ref.~\cite{Panero:2004zq} for the four-dimensional case: following the notations of ref.~\cite{Caselle:2014eka}, the partition function defined in eq.~(\ref{lattice_partition_function}) can be rewritten as
\begin{equation}
Z = \sum_{ \{ {}^\star s \in \Z \} } \prod_{\mbox{\tiny{bonds}}} I_{|d {}^\star s|} (\beta),
\end{equation}
where $I_\nu(z)$ is the modified Bessel function of the first kind of order $\nu$, the product is taken over the bonds of the dual cubic lattice, and $d {}^\star s$ denotes the difference  of the integer-valued ${}^\star s$ variables on the two sites at the ends of each bond.

The Monte~Carlo evolution of the system is dictated by the evolution of the configuration of the frustrated plaquettes (those where $^\star n \neq 0$); to enhance the numerical precision of the simulation results, our algorithm features a hierarchical sequence of nested updates~\cite{Caselle:2002ah}.

Given a $Q\bar{Q}$ pair of static sources at a distance $R$ from each other in the original theory, the associated two-point correlation function of Polyakov loops can then be rewritten as
\begin{equation}
\langle P^\star (R) P(0) \rangle = \frac{Z_R}{Z}  = \frac{1}{Z}\sum_{ \{ {}^\star s \in \Z \} } \prod_{\mbox{\tiny{bonds}}} I_{|d {}^\star s+ {}^\star n|} (\beta),
\end{equation}
where ${}^\star n$ denotes an integer-valued 1-form which is non-vanishing on a set of bonds, that are dual to the plaquettes tiling a surface bounded by the Polyakov loops. Finally, the connected correlation function $e_l (x_t)$ between the Polyakov loops and the electric field component in the direction parallel to the $Q\bar{Q}$ axis at the point $x_t$ can be written as
\begin{equation}
\label{el}
e_l (x_t) = \frac{\langle P^\star (R) P(0) E_l (x_t) \rangle}{\langle P^\star (R) P(0) \rangle} - \langle E_l (x_t) \rangle = \frac{\langle\dd^\star l + ^\star n\rangle}{\sqrt{\beta}}.
\end{equation}
The squared width of the flux tube is then computed as the second moment of eq.~(\ref{el}):
\begin{equation}
\label{squared_width}
w^2 = \frac{\sum_{|x_t| \le x_{\text{max}}} x_t^2 e_l(x_t) }{\sum_{|x_t| \le x_{\text{max}}} e_l(x_t)},
\end{equation}
where $x_{\text{max}}$ is fixed by the requirement that the contributions to the sums from points at $|x_t| > x_{\text{max}}$ are not significant, within the numerical precision of our results.\footnote{Although, in principle, the discretized numerical integrations involved in eq.~(\ref{squared_width}) could be carried out using some improved computational technique (see, for example, ref.~\cite{Caselle:2007yc} and the references therein), the corresponding reduction in systematic uncertainties would have a negligible impact on the error budget of our results.}

We computed $e_l$ and $w^2$ for $\beta=1.7$, $2.0$, $2.2$ and $2.4$. Additional information on our simulations is reported in table~\ref{tab:simulsetting}.

\begin{table}[!htb]
\centering
\begin{tabular}{|c|c|c|c|}
\hline
$\beta$  &$\sigma a^2$  & $m_0 a$ & $L$, $N_t$ \\
\hline
$1.7$ & $0.122764(2)$ & $0.889(4)$   & $64$ \\
$2.0$ & $0.049364(2)$ & $0.449(4)$   & $64$ \\
$2.2$ & $0.027322(2)$ & $0.27(1)$   & $96$ \\
$2.4$ & $0.015456(7)$ & $0.165(10)$ & $96$ \\
\hline
\end{tabular}
\caption{Information on the parameters of our simulations.}
\label{tab:simulsetting}
\end{table}

\section{Results}
\label{sec:results}

In this section, we present our numerical results (and their analysis) for the squared width of the flux tube in subsection~\ref{subsec:width}, and for the flux-tube profile in subsection~\ref{subsec:profile}.

\subsection{Squared width of the flux tube}
\label{subsec:width}

In table~\ref{tab:w2-vs-R}, we report our results for the squared width of the flux tube on the mid-plane between the Polyakov loops, for different values of the spatial separation between the charges (and at different values of the coupling).

\begin{table}[ht]
\centering
\begin{tabular}{|c|c|c|c|c|}
\hline
& \multicolumn{4}{|c|}{$w^2/a^2$} \\
\hline
$R/a$  & $\beta=1.7$ & $\beta=2.0$ & $\beta=2.2$ & $\beta=2.4$\\
\hline
 $4$ &    --      & $8.59(21)$  & $14.26(28)$ & $22.4(9)$   \\
 $6$ & $4.00(10)$ & $10.46(28)$ & $19.1(5)$   & $31.8(1.0)$ \\
 $8$ & $4.32(6)$  & $12.20(24)$ & $22.1(6)$   & $42.1(1.7)$ \\
$10$ & $4.74(6)$  & $13.75(27)$ & $25.8(7)$   & $47.2(1.4)$ \\
$12$ & $4.98(6)$  & $14.72(31)$ & $28.4(7)$   & $49.2(3.3)$ \\
$14$ & $5.23(6)$  & $15.49(27)$ & $30.6(1.0)$ & $55.4(2.8)$ \\
$16$ & $5.43(6)$  & $16.52(31)$ & $33.2(9)$   & $67.2(3.2)$ \\
$18$ & $5.63(8)$  & $16.9(4)$   & $35.3(1.1)$ & $65.1(3.3)$ \\
$20$ & $5.78(8)$  & $17.1(4)$   & $35.4(1.0)$ & $64.6(3.3)$ \\
$22$ & $5.81(6)$  & $18.4(4)$   & $38.1(9)$   & $71.5(3.2)$ \\
$24$ &    --      &    --       & $37.1(8)$   & $76.6(3.5)$ \\
$26$ &    --      &    --       & $38.9(1.0)$ & $75.3(3.3)$ \\
$28$ &    --      &    --       & $41.6(1.2)$ & $76.7(3.4)$ \\
$30$ &    --      &    --       &    --       & $82.5(2.0)$ \\
$32$ &    --      &    --       &    --       & $85.3(1.9)$ \\
$34$ &    --      &    --       &    --       & $86.9(1.7)$ \\
$36$ &    --      &    --       &    --       & $90.4(1.9)$ \\
$38$ &    --      &    --       &    --       & $86.7(2.0)$ \\
$40$ &    --      &    --       &    --       & $91.8(2.1)$ \\
$42$ &    --      &    --       &    --       & $90.7(2.4)$ \\
\hline                                                   
\end{tabular}
\caption{Square width $w^2$ of the flux tube at different values of $\beta=1/(ae^2)$, as a function of the distance $R$ between the static sources.}
\label{tab:w2-vs-R}
\end{table}

A glance at table~\ref{tab:w2-vs-R} immediately reveals that $w^2/a^2$ is certainly not constant when $R/a$ grows. Thus, we fit our data to the following form: 
\begin{equation}
\label{f1new}
w^2=\frac{1}{2\pi s}\log(R/R_0),
\end{equation} using $s$ and $R_0$ as fitting parameters. If the Nambu-Got\={o} string provided a correct low-energy model of flux tubes in this theory, then eq.~(\ref{f1new}) should describe the numerical data for $R \gtrsim 1/\sqrt{\sigma}$, and $s$ should coincide with $\sigma$, the string tension extracted from the linear behavior of the interquark potential at large distances (for which we use the results of ref.~\cite{Caselle:2014eka}).

\begin{figure}[!htpb]
\centering{
\includegraphics[width=0.48\textwidth]{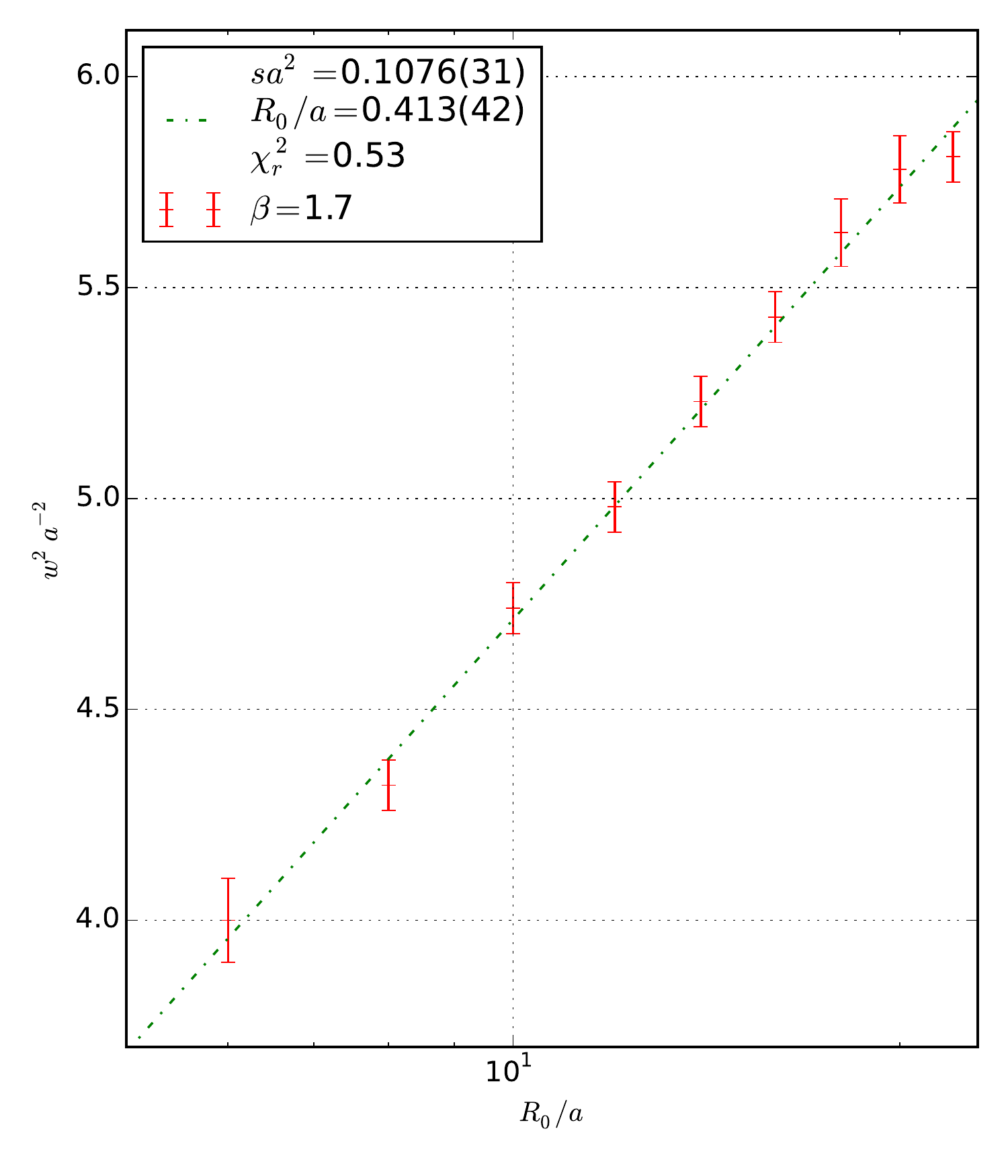}\hfill
\includegraphics[width=0.48\textwidth]{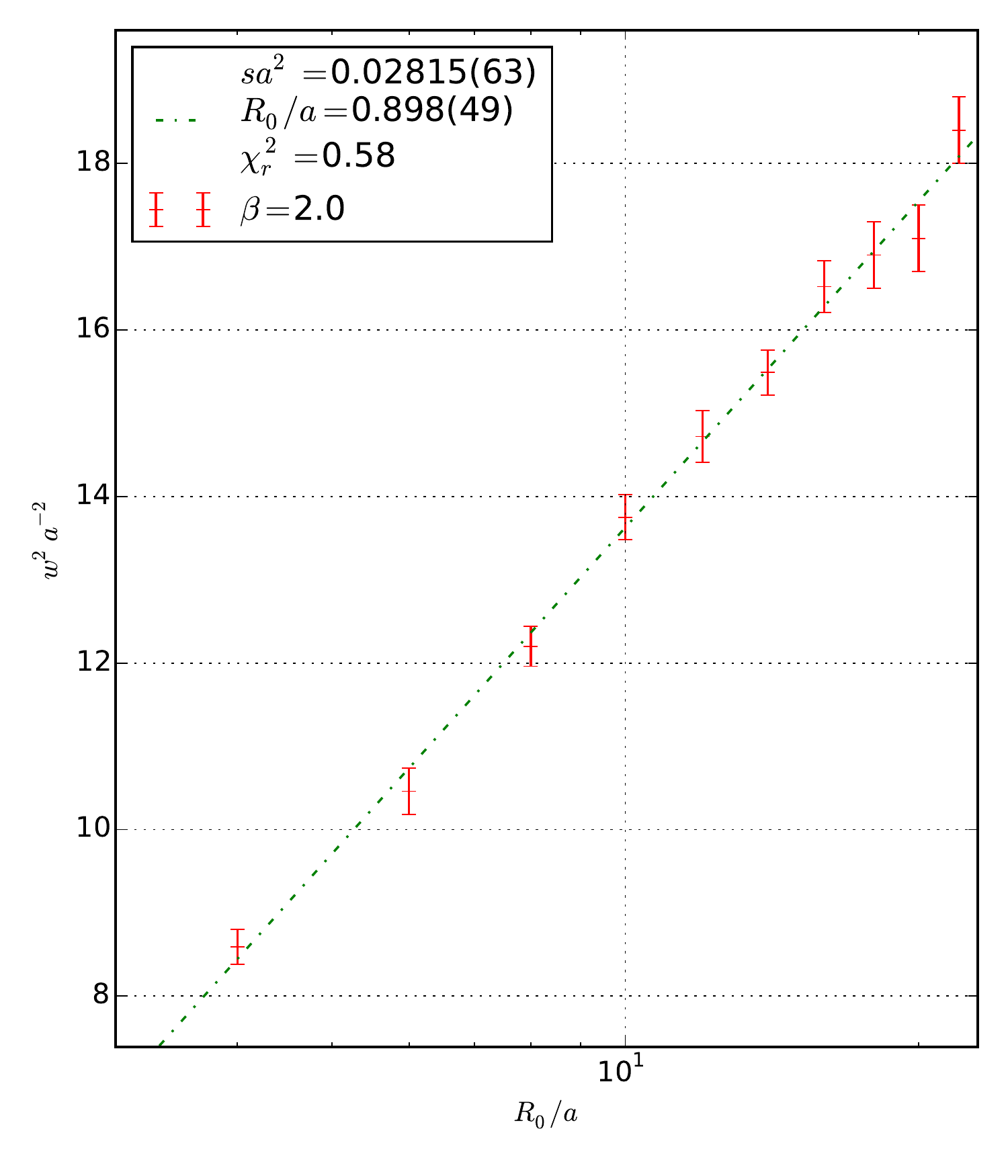}
}
\vspace{4mm}
\centering{
\includegraphics[width=0.48\textwidth]{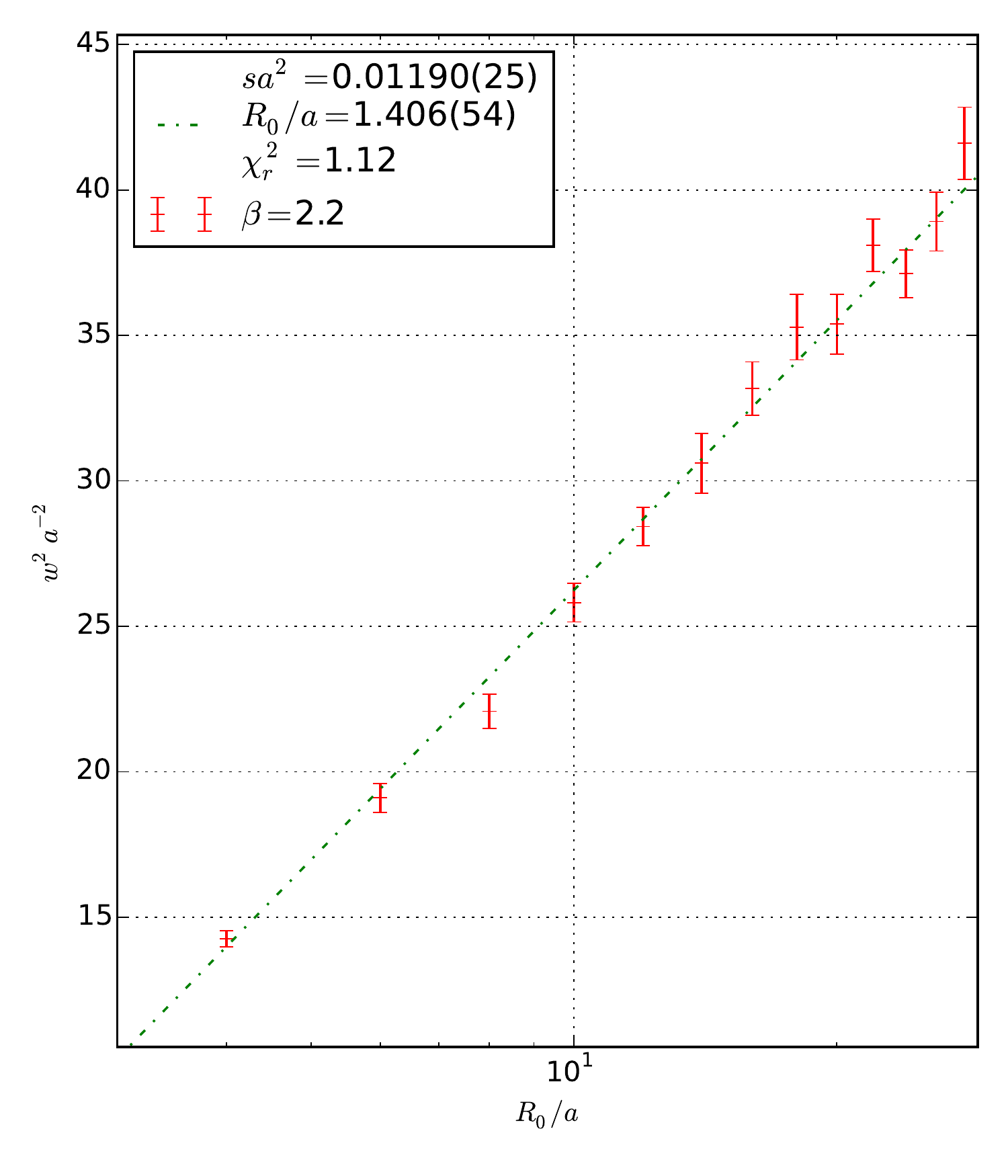}\hfill
\includegraphics[width=0.48\textwidth]{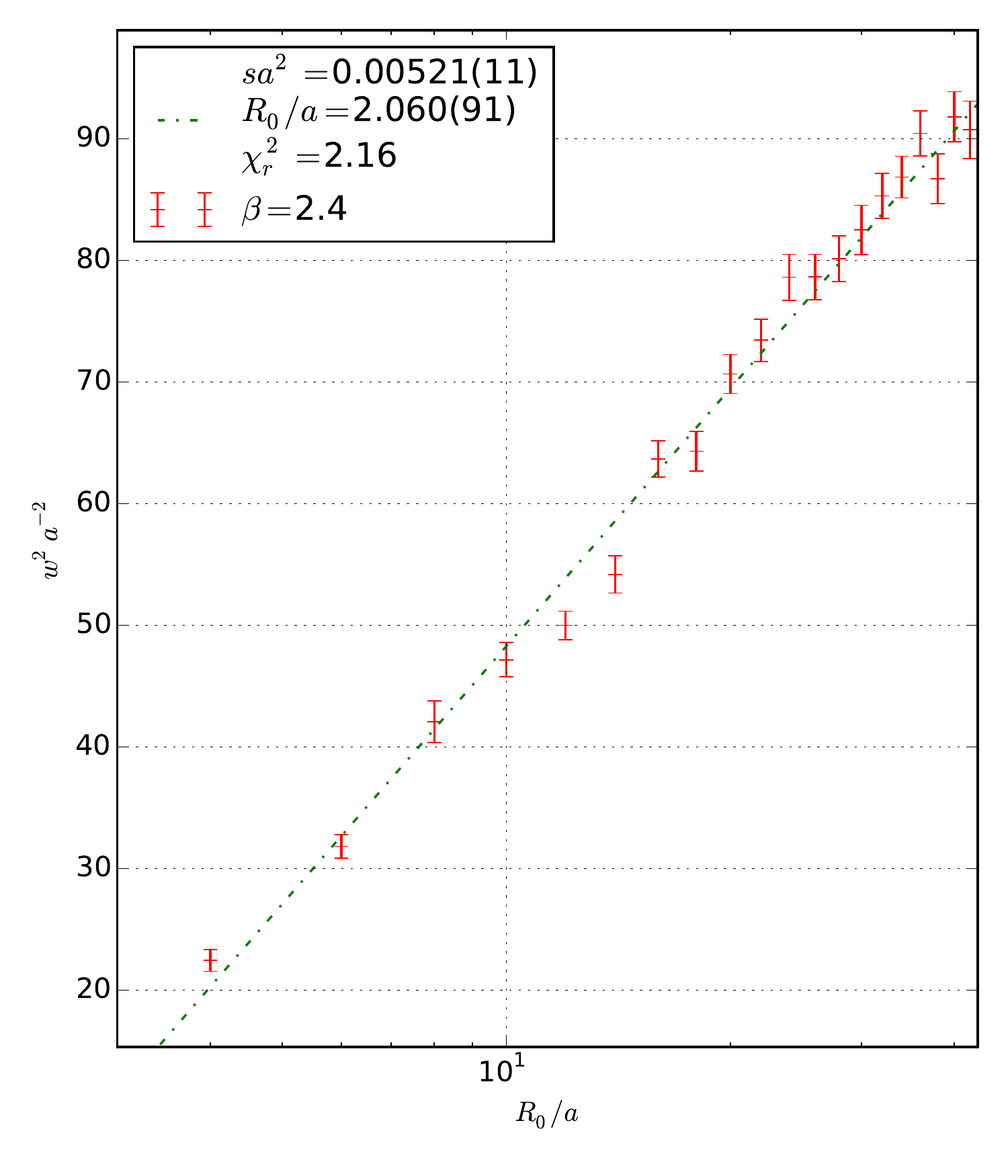}
}
\caption{Results for $w^2/a^2$, as a function of $R/a$, from our simulations at $\beta=1.7$~(top left panel), at $\beta=2.0$~(top right panel), at $\beta=2.2$~(bottom left panel), and at $\beta=2.4$~(bottom right panel). Note that the horizontal axis is displayed in logarithmic scale. The dashed lines are the curves obtained from two-parameter fits to eq.~(\ref{f1new}), for the values of $sa^2$ and $R_0/a$ displayed in the legend box of each plot.}
\label{fig:log}
\end{figure}

Starting with a fit to the whole range of $R$, we progressively discarded the data at the smallest $R$, until the reduced $\chi^2$ of the fit ($\redchisq$) became close to $1$: the results of these fits are reported in table~\ref{tab:w2-fitlog}, and plotted (along with the Monte~Carlo data) in figure~\ref{fig:log}; note that the horizontal axis of this figure is in logarithmic scale.

\begin{table}[!htpb]
\centering
\begin{tabular}{|c|c|c|c|c|c|}
\hline
$\beta$ & $R_{\text{min}}/a$ & $s a^2$ & $R_0/a$ & $\redchisq$ &  d.o.f.\\
\hline
 $1.7$ &  $ 6 $ & $ 0.1076(31)  $ & $ 0.413(42) $ & $ 0.53 $ & $ 7  $ \\
 $2.0$ &  $ 4 $ & $ 0.02815(63) $ & $ 0.898(49) $ & $ 0.58 $ & $ 8  $ \\
 $2.2$ &  $ 4 $ & $ 0.01190(25) $ & $ 1.406(54) $ & $ 1.12 $ & $ 11 $ \\
 $2.4$ &  $ 4 $ & $ 0.00521(11) $ & $ 2.060(91) $ & $ 2.16 $ & $ 18 $ \\
\hline
\end{tabular}
\caption{Results of the logarithmic fits of $w^2$ to eq.~(\ref{f1new}), at the different values of $\beta$ that we studies (reported in the first column). The second column shows the minimal value of $R/a$ included in the fit, while the third and the fourth column show the fitted parameters. The reduced $\chi^2$ of the fit and the number of degrees of freedom (d.o.f.) 
are shown in the fifth and in the sixth column, respectively.}
\label{tab:w2-fitlog}
\end{table}

As one can easily see, eq.~(\ref{f1new}) gives a good description of the data for a wide range of values of $R$. However, table~\ref{tab:w2-fitlog} reveals that the fitted value of $s$ is incompatible with the string tension, in contrast with the prediction from the effective string picture (in the Gau{\ss}ian approximation that we are considering).

\begin{figure}[!htpb]
\centerline{\includegraphics[width=0.7\textwidth]{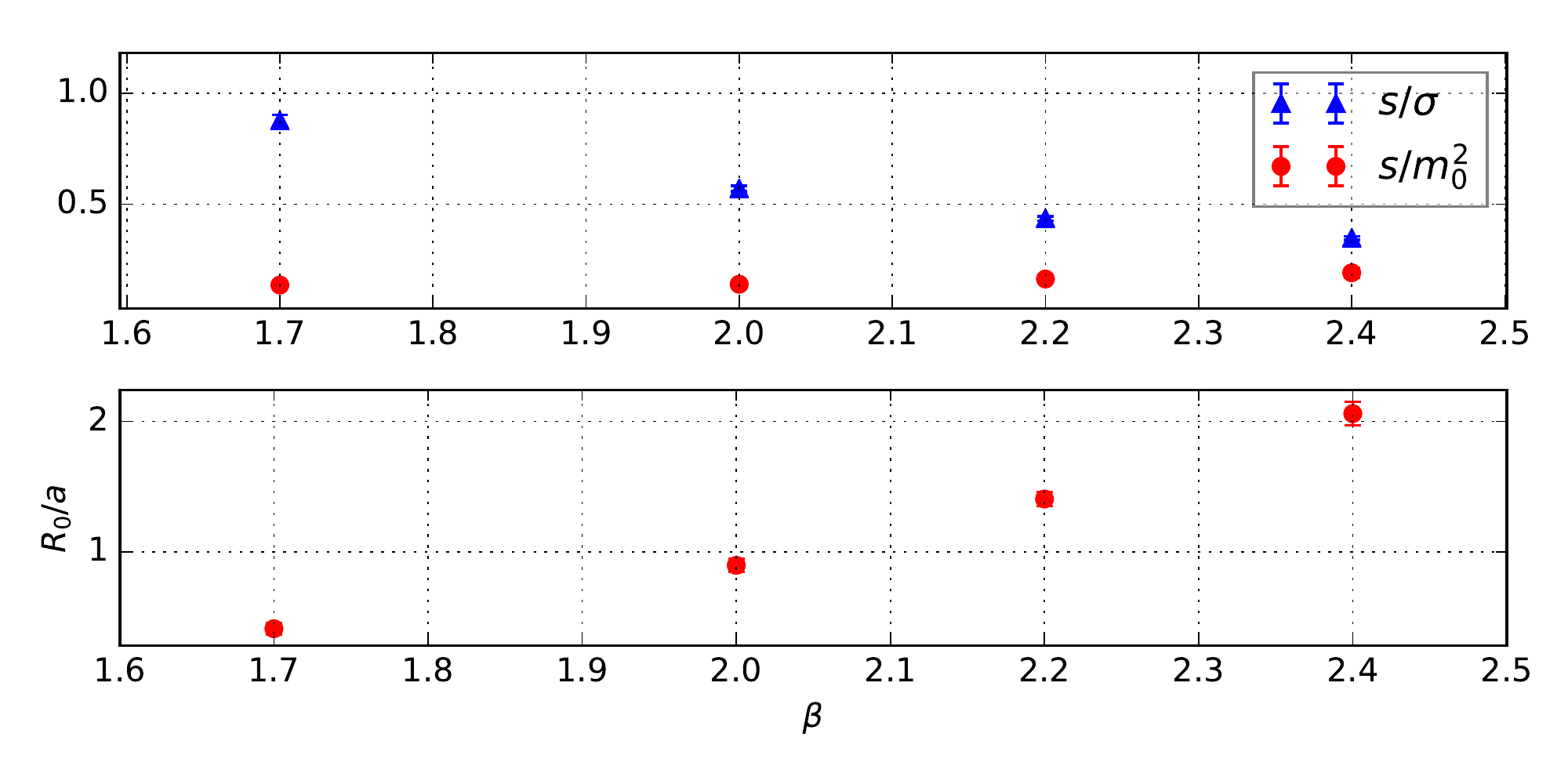}}
\caption{Top panel: The ratio $s/\sigma$ between the parameter $s$ obtained from the fit to~eq.\ref{f1new} and the string tension $\sigma$ determined from fits to the interquark potential; for comparison, the plot also shows the ratio between $s$ and the square of the glueball mass $m_0$. Bottom panel: The values of $R_0/a$ obtained from the fit.}
\label{fig:srcvsbeta}
\end{figure}

The deviation of $s$ from the string tension can be better appreciated in table~\ref{tab:s_sigma_m0square}, in which we also compared $s$ with (the square of) the other physical scale of the theory, i.e. the mass gap. These results are also shown in the top panel of figure~\ref{fig:srcvsbeta} (while the bottom panel of the same figure shows how the other fitted parameter, $R_0/a$, depends on $\beta$).

\begin{table}[!htpb]
\centering
\begin{tabular}{|c|c|c|}
\hline
$\beta$ & $s/\sigma$ & $s/m_0^2$ \\
\hline
$ 1.7 $ & $ 0.876(25) $ & $ 0.136(4) $ \\
$ 2.0 $ & $ 0.570(13) $ & $ 0.140(4) $ \\
$ 2.2 $ & $ 0.436(9) $ & $ 0.163(13) $ \\
$ 2.4 $ & $ 0.347(8) $ & $ 0.191(21) $ \\
\hline
\end{tabular}
\caption{Comparison of the $s$ parameter, obtained from fits of $w^2$ to eq.~(\ref{f1new}) at the different $\beta$ values (listed in the first column), to the corresponding values of the string tension $\sigma$ (second column) and to the squared mass gap in lattice units $m_0^2$ (third column).}
\label{tab:s_sigma_m0square}
\end{table}

Moreover, our results at large $\beta$ show that the fits to a purely logarithmic form, eq.~(\ref{f1new}), hold also for distances much shorter than the length scale, below which the effective string picture is expected to break down, namely $1/\sqrt{\sigma}$. For example, at $\beta=2.4$ this quantity corresponds to more than eight lattice spacings, while all data (starting from $R\ge 4a$) can be successfully fitted to eq.~(\ref{f1new}).

We also verified that the deviations from the Nambu-Got{\={o}} prediction cannot be explained by the next-to-leading order effects discussed in refs.~\cite{Gliozzi:2010zv, Gliozzi:2010zt}. In particular, these effects play a negligible r\^{o}le at large distances (and cannot account for the strong discrepancy between $s$ and $\sigma$ that the numerical data reveal); their impact becomes non-negligible at intermediate distances, but their form is clearly incompatible with the simple logarithmic behavior exhibited by our Monte~Carlo results.

As an alternative explanation for the mismatch between the EST model and the simulation results, one could imagine that the deviations from the Nambu-Got{\={o}} prediction be due to the presence of extrinsic-curvature terms in the effective string action~\cite{Caselle:2014eka}. However, a complete analytical derivation of the contributions to $w^2$ from such terms is non-trivial, and lies clearly beyond the scope of the present work.

These results show that the Nambu-Got{\={o}} string model \emph{does not} provide an accurate description of flux tubes in this theory.

Comparing the fit results in table~\ref{tab:w2-fitlog} with the values for $\sigma a^2$ in table~\ref{tab:simulsetting}, it is also interesting to note that the value of $R_0$ extracted from the fits does not scale like $1/\sqrt{\sigma}$: as the lattice spacing is decreased, the dimensionless product $R_0\sqrt{\sigma}$ grows from $0.145(15)$ for $\beta=1.7$ to $0.256(11)$ at $\beta=2.4$. Clearly, this is at odds with the expectation that $R_0$ should tend to a well-defined constant value in the continuum limit (as is the case in non-Abelian gauge theories in 4D). Instead, our data give a clear indication that, in this theory, the minimal tube length, below which the Nambu-Got{\={o}} string description breaks down, is monotonously \emph{increasing} to larger and larger values for $a \to 0$.

\begin{figure}[!htpb]
\centering{
\includegraphics[height=0.24\textheight]{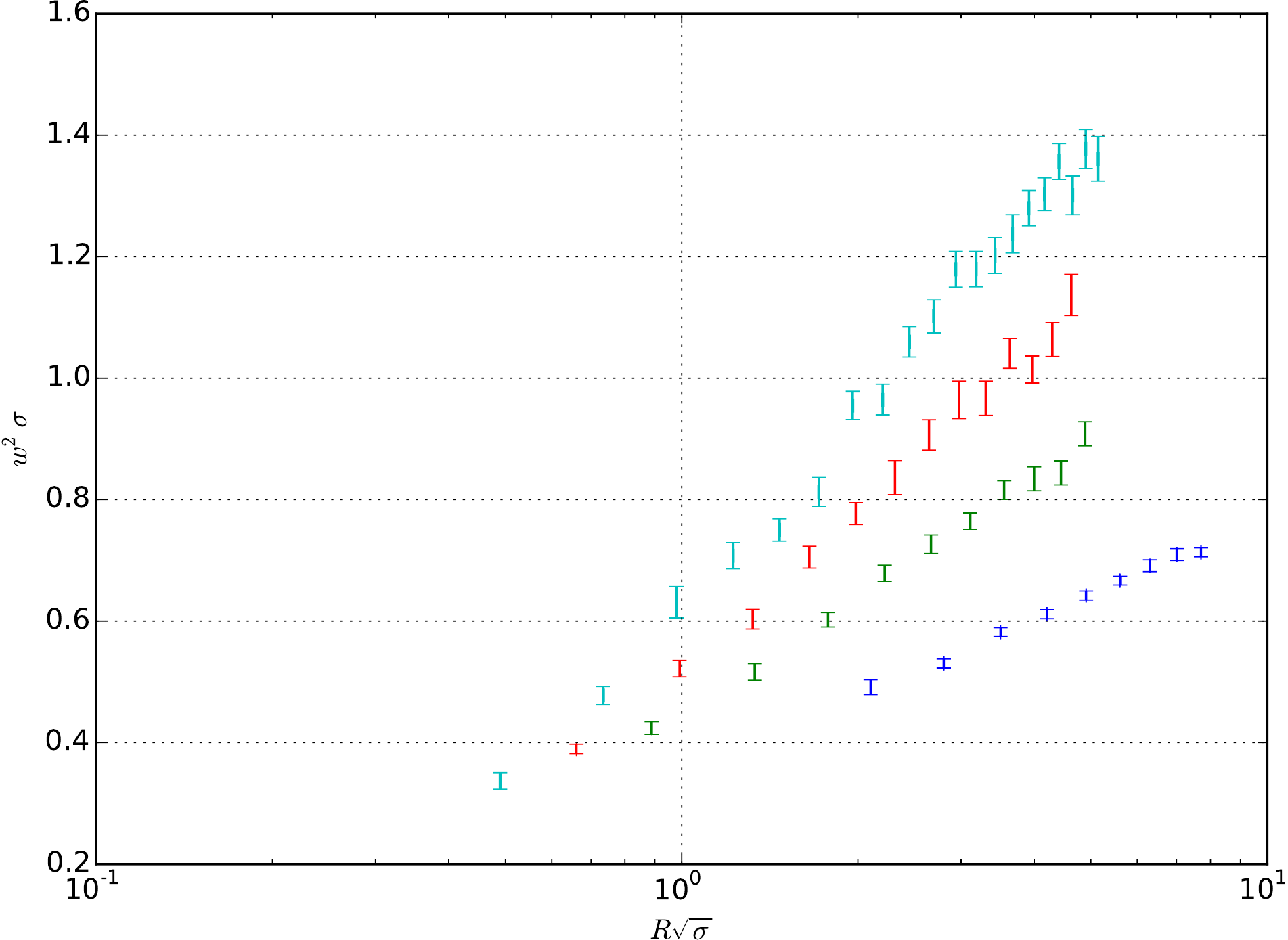}\hfill
\includegraphics[height=0.24\textheight]{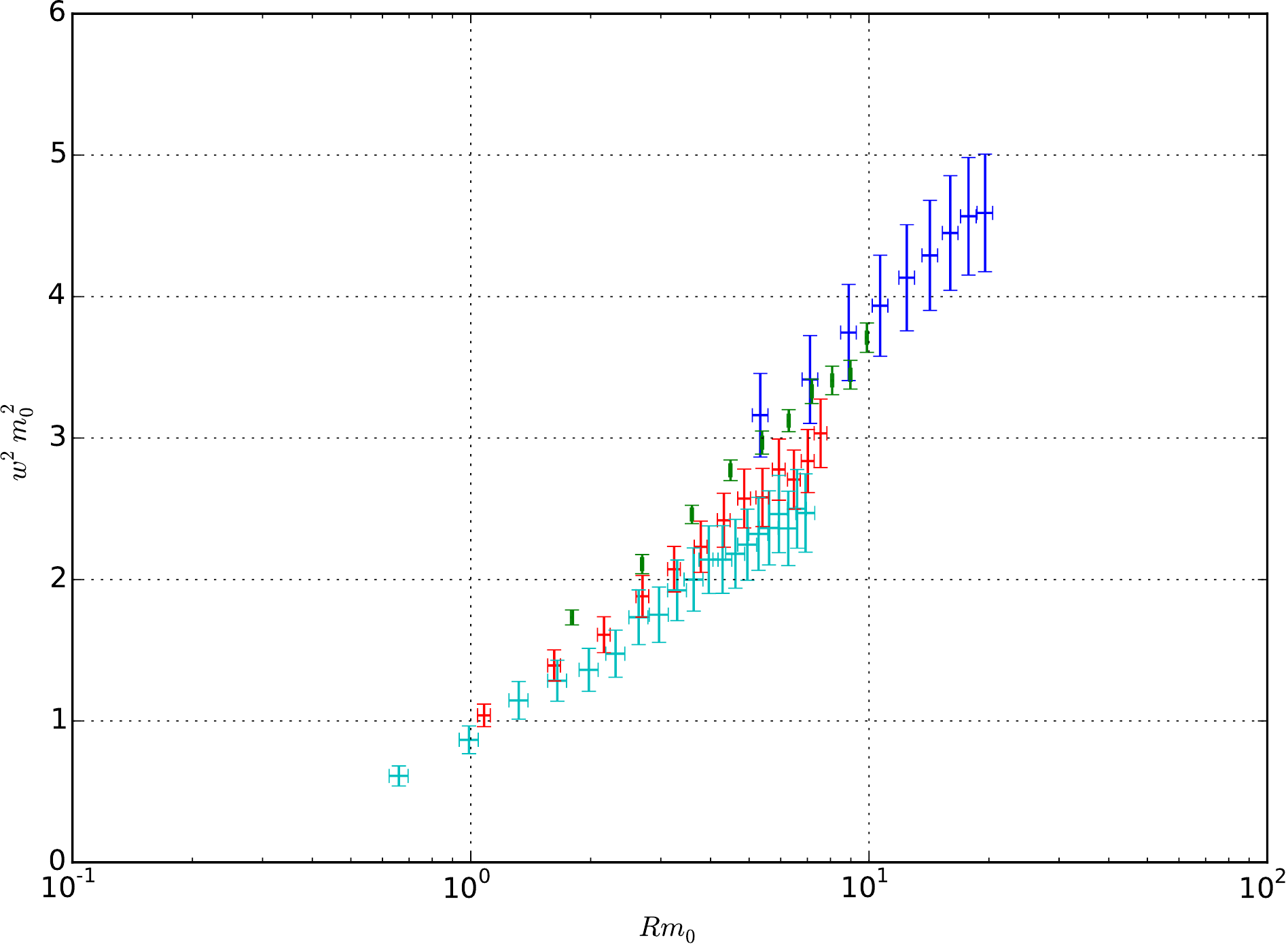}
}
\caption{The lattice results for $w^2$ as a function of $R$ at different values of $\beta$ (displayed by symbols of different colors: dark blue for $\beta=1.7$, dark green for $\beta=2.0$, red for $\beta=2.2$ and cyan for $\beta=2.4$) do not fall on a common curve when plotted in units of $1/\sqrt{\sigma}$ (l.h.s. panel), while they exhibit a nearly perfect collapse to the same curve, when expressed in units of the inverse of the mass gap $1/m_0$ (r.h.s. panel). Note the logarithmic scale for the horizontal axis.}
\label{fig:collapse_of_data}
\end{figure}

Another observation, indicating that $1/\sqrt{\sigma}$ is \emph{not} the ``natural'' physical length scale of this model is shown in fig.~\ref{fig:collapse_of_data}: our lattice results at different lattice spacing do not fall on a universal curve when expressed in units of $1/\sqrt{\sigma}$ (left-hand-side panel), but they do, when they are plotted in units of $1/m_0$ (right-hand-side panel).

In order to get a better understanding of the dynamics of the flux tube, it is useful to directly study its profile. Thanks to the dual algorithm, it is possible to obtain very precise results for this quantity for a broad range of values of $R$, and at large distances from the axis between the sources. The results of this study are presented in the following subsection.

\subsection{Flux-tube profile}
\label{subsec:profile}

The various models of the confining flux tube yield very different predictions for its ``shape''---i.e. for the dependence of the field strength on the transverse separation from the source-source axis. As mentioned in section~\ref{sec:introduction}, two of the most commonly used models are the one describing the flux tube as a fluctuating bosonic string, and the one based on the interpretation of the vacuum of a confining gauge theory as a dual (color-)superconductor, in which flux tubes are interpreted as ``Abrikosov vortices''. In this subsection, we will compare their predictions with our results for compact $\U(1)$ lattice gauge theory in 3D.

\subsubsection{Comparison with the bosonic-string model}
\label{subsubsec:bosonic-string}

At the leading order in an expansion around the classical, straight-tube configuration, the effective string model predicts that the probability $p$ of finding the flux tube at a transverse displacement $x_t$ from the axis of the color sources decreases exponentially:
\begin{equation}
\label{Gaussian}
p(x_t) = C_0 \exp \left(-x_t^2/ \delta^2\right),
\end{equation}
where $C_0$ is independent of $x_t$, and $\delta^2$ is defined in terms of a sum over the allowed vibration modes of the string~\cite{Luscher:1980iy}. This relation implies eq.~(\ref{f1}) (with $w^2=\delta^2$). In order to test this prediction, we attempted to fit our Monte~Carlo results for the ``tails'' of the $e_l(x_t)$ profile with a Gau{\ss}ian of the form of eq.~(\ref{Gaussian}). More precisely, we tried a Gau{\ss}ian fit of $e_l(x_t)$ for distances from the sources' axis larger than a certain value $x_t^c$. To fix the latter, we started from $x_t^c=0$ and progressively increased its value, until both the following conditions were met:
\begin{enumerate}
\item the $\redchisq$ was close to unity, and
\item by further increasing $x_t^c$, the fitted parameter values did not change in a statistically significant way. 
\end{enumerate}
An example of the results obtained from this analysis (for $\beta=2.0$ and interquark distance $R=10a$) is reported in table~\ref{tab:fit-gauss-ko}: it shows that there exists no distance range in the $e_l(x_t)$ tails, that can be described by a pure Gau{\ss}ian. We reached the same conclusions also for all the other data sets (corresponding to different values of $\beta$ and/or $R$) that we studied: for this lattice theory, the inadequacy of a Gau{\ss}ian description for the tails of the flux tube profile seems to be generic.

\begin{table}[ht]
\centering
\begin{tabular}{|c|c|c|c|}
\hline
$x_t^c/a$ & $C_0$ & $a^2/\delta^2$ & $\redchisq$ \\
\hline
$0 $&$ 0.00943(4) $&$ 0.1903(4) $&$ 11.64 $ \\
$1 $&$ 0.00934(4) $&$ 0.1895(5) $&$ 11.19 $ \\
$2 $&$ 0.00905(5) $&$ 0.1872(5) $&$ 9.68 $  \\
$3 $&$ 0.00873(5) $&$ 0.1846(6) $&$ 8.62 $  \\
$4 $&$ 0.00813(7) $&$ 0.1801(6) $&$ 6.24 $  \\
$5 $&$ 0.00756(8) $&$ 0.1760(8) $&$ 5.1 $   \\
$6 $&$ 0.00696(11) $&$ 0.1717(9) $&$ 4.31 $ \\
$7 $&$ 0.00604(13) $&$ 0.1652(11) $&$ 3.0 $ \\
$8 $&$ 0.00532(16) $&$ 0.1600(14) $&$ 2.47$ \\
$9 $&$ 0.00451(19) $&$ 0.1539(17) $&$ 2.05$ \\
\hline 
\end{tabular}
\caption{Results for the fit to eq.~(\ref{Gaussian}) of our data for the flux-tube profile at $\beta=2.0$ and for $R=10a$.}
\label{tab:fit-gauss-ko}
\end{table}

\subsubsection{Comparison with the dual-superconductor model}
\label{subsubsec:dual_superconductor}

\begin{figure}[!htpb]
\centerline{\includegraphics[width=0.7\textwidth]{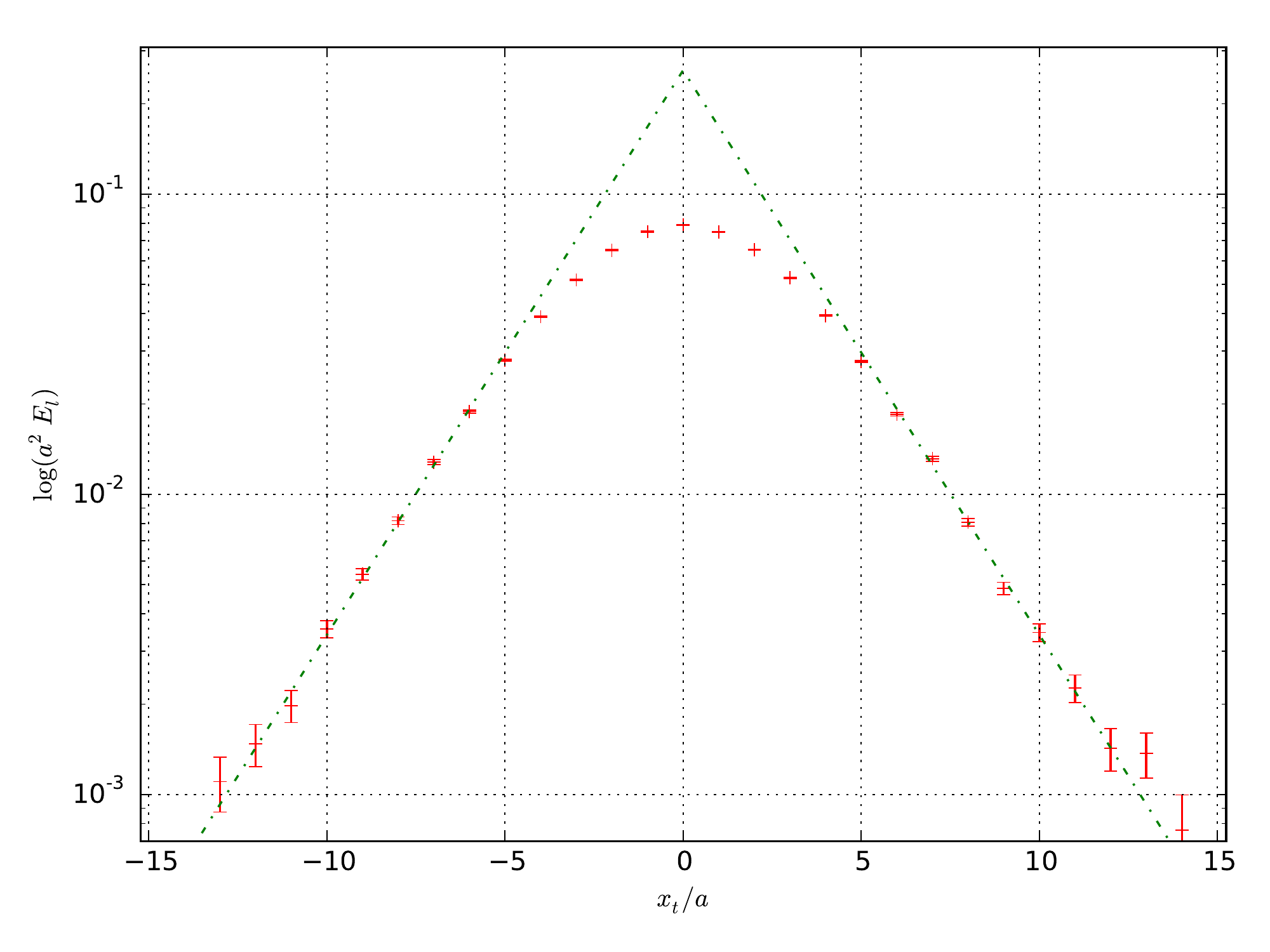}}
\caption{The transverse profile of the flux tube, for $\beta=2.0$ and $R=16a$. Note the logarithmic scale for the vertical axis. In this plot, the exponential decay of the tails of $e_l(x_t)$ manifests itself in the linear behavior observed for $\left| x_t \right| \ge 8a$: the green dash-dotted line is the result of the fit to eq.~(\ref{tab:fit-exp}) in this range.}
\label{fig:logElvsz-2.0-16a}
\end{figure}

The dual-superconductor model predicts that, far from the core of the dual Abrikosov vortex sheet, a uniform density of magnetic color monopoles leads to a 
\begin{equation}
e_l(x_t) = \Phi m_v^2 \exp\left(- m_v \left| x_t \right| \right),
\label{dualsup}
\end{equation}
where $\Phi$ is the total flux carried by the tube and $m_v$ is the mass of the gauge field (which is related to the London penetration length $\mu$ as $m_v=1/\mu$). We fitted our Monte~Carlo results for the tails of $e_l(x_t)$ to eq.~(\ref{dualsup}), using the same method described above, and found that this functional form successfully describes the tails of the profiles, and that the fitted parameters are robust, in the sense that they do not change in a statistically significant way, if $x_t^c$ is increased to values larger than the minimal distance for which a $\redchisq \simeq 1$ is obtained. Examples of parameters obtained from this analysis are reported in table~\ref{tab:fit-exp} (for the data at $\beta=2.0$ and interquark separation $R=10a$) and in table~\ref{tab:muvsxtc-2.0-16} (for $\beta=2.0$ and $R=16a$: the data of this set are plotted in figure~\ref{fig:logElvsz-2.0-16a}), but the result is generic.

\begin{table}[!htpb]
\centering
\begin{tabular}{|c|c|c|c|c|c|c|c|}
\hline
$x_t^c/a$ & $\Phi$ & $a m_v$ & $\redchisq$ \\
\hline
$1 $ & $7.4(2)$   & $0.323(8)$ & $93.5$ \\
$2 $ & $6.9(1)$   & $0.368(6)$ & $21.5$ \\
$3 $ & $6.86(4)$  & $0.405(4)$ & $3.86$ \\
$4 $ & $6.99(3)$  & $0.428(4)$ & $1.48$ \\
$5 $ & $7.21(6)$  & $0.444(5)$ & $0.96$ \\
$6 $ & $7.25(13)$ & $0.447(7)$ & $0.96$ \\
$7 $ & $7.11(28)$ & $0.44(1)$  & $1.03$ \\
$8 $ & $7.3(6)$   & $0.44(2)$  & $1.06$ \\
$9 $ & $6.9(1.0)$ & $0.44(3)$  & $1.12$ \\
$10$ & $6.5(1.9)$ & $0.43(5)$  & $1.18$ \\
\hline 
\end{tabular}
\caption{Results of the fits to eq.~(\ref{dualsup}) for the Monte~Carlo data obtained at $\beta=2.0$ and $R=10a$.}
\label{tab:fit-exp}
\end{table}

\begin{table}[!htpb]
\centering
\begin{tabular}{|c|c|c|c|c|c|}
\hline
$x_t^c/a$ & $\Phi$ & $a m_v$ & $\redchisq$ \\
\hline
 $1$ & $1.34(6)$   & $0.284(9)$  & $115.19$ \\
 $2$ & $1.246(29)$ & $0.325(7)$  & $35.61$  \\
 $3$ & $1.224(14)$ & $0.360(5)$  & $10.4$   \\
 $4$ & $1.249(7)$  & $0.390(4)$  & $2.96$   \\
 $5$ & $1.297(9)$  & $0.411(5)$  & $1.46$   \\
 $6$ & $1.336(21)$ & $0.422(7)$  & $1.33$   \\
 $7$ & $1.47(5)$   & $0.446(9)$  & $1.0$    \\
 $8$ & $1.38(8)$   & $0.433(14)$ & $0.99$   \\
 $9$ & $1.32(15)$  & $0.426(22)$ & $1.04$   \\
$10$ & $1.51(32)$  & $0.445(34)$ & $0.99$   \\
$11$ & $1.4(5)$    & $0.43(5)$   & $1.06$   \\
$12$ & $2.3(1.8)$  & $0.49(9)$   & $1.08$   \\
\hline 
\end{tabular}
\caption{Same as in table~\ref{tab:fit-exp}, but for data at $\beta=2.0$ and $R=16a$.}
\label{tab:muvsxtc-2.0-16}
\end{table}

We conclude that the prediction of the dual-superconductor model, eq.~(\ref{dualsup}), provides a correct description of the large-$x_t$ behavior of our Monte~Carlo data. In addition, we also found that $m_v$, which should be a property of the confining medium (not of the particular configuration of the probe color sources), is indeed independent of the charge separation $R$ (in fact, the dependence on $R$ is encoded solely in $x_t^c$ and in the amplitude $\Phi$). When expressed in units of the glueball mass $m_0$, $m_v$ takes a value independent of $\beta$, and compatible with $1$ (see table~\ref{tab:muvsbeta}): the London penetration length of the model is thus consistent with the inverse of the mass gap.

\begin{table}[ht]
\centering
\begin{tabular}{|c|c|c|c|c|}
\hline
$\beta$ & $m_v/m_0$ \\
\hline
$1.7$&$0.99(5)$\\
$2.0$&$1.012(16)$\\
$2.2$&$1.03(4)$\\
$2.4$&$1.06(6)$\\
\hline
\end{tabular}
\caption{Values of the inverse London penetration depth at varying $\beta$, in units of the lightest glueball mass, obtained from fits to eq.~(\ref{dualsup}).}
\label{tab:muvsbeta}
\end{table}

\section{Discussion and concluding remarks}
\label{sec:discussion_and_concluding_remarks}

The results that we presented in section~\ref{sec:results} lead one to conclude that, although in this gauge theory the broadening of the confining flux tube is consistent with a logarithmic dependence on the tube length, as predicted by a bosonic-string model~\cite{Luscher:1980iy}, the amplitude of the tube width is not. In addition, the logarithmic dependence on the length of the flux tube persists also for tubes shorter than the characteristic length scale $1/\sqrt{\sigma}$, below which the Nambu-Got{\={o}} string prediction is not expected to hold anymore. Similarly, the values obtained for $R_0$---that is the length scale, below which eq.~(\ref{f1}) necessarily breaks down\footnote{Obviously, eq.~(\ref{f1}) is meaningful only for $R$ larger than $R_0$; in the opposite range, the logarithm turns negative, and eq.~(\ref{f1}) would then imply the unphysical result $w^2<0$.}---indicate that, when the lattice spacing $a$ is decreased to sufficiently small values, the Nambu-Got{\={o}} string model eventually fails to describe the width of any flux tube of fixed length. This discrepancy between the string model and Monte~Carlo data cannot be accommodated invoking higher-order terms derived from the Nambu-Got\={o} action.

A closer inspection of the flux-tube profile reveals that its decay at large distance from the interquark axis is clearly incompatible with the leading-order EST prediction, while it fully agrees with the predictions of the dual-superconductor model. In particular, the London penetration length is equal to the inverse of the mass of the lightest glueball; remarkably, this holds at all of the $\beta$ values considered in this work.

These results reinforce the conclusions of our recent study of the interquark potential in this model~\cite{Caselle:2014eka}, where we found that, as $\beta$ is increased toward the continuum limit, the behavior of the flux tube shows larger and larger deviations from the prediction of the Nambu-Got{\={o}} model. The flux-tube width, that we investigated in the present work, exhibits a similar pattern: for example, fitting our Monte~Carlo results for $w^2$ at $\beta=1.7$ to eq.~(\ref{f1new}), we obtained a value for $s$ (in lattice units) that is almost compatible with $\sigma$, but when $\beta$ is increased the two quantities become more and more  divergent from each other. In ref.~\cite{Caselle:2014eka}, we also observed that, as $\beta$ is increased, the minimal distance (in \emph{physical} units), at which the Nambu-Got{\={o}} model describes the data well, is pushed to larger and large values. Here, the increase of $R_0\sqrt{\sigma}$ with $\beta$, discussed in subsection~\ref{subsec:profile}, is another facet of the same effect.

In ref.~\cite{Caselle:2014eka}, we put forward the hypothesis that the deviations from the Nambu-Got{\={o}} model signal the presence of an extrinsic-curvature term in the effective string action. More precisely, we showed that the results for the confining potential are compatible with such term, if it appears with a coefficient $\alpha$ proportional to $1/m_0$, so that the $\alpha/\sigma$ ratio diverges like $1/m_0^2$ in the continuum limit. The present analysis supports this picture, as the shape of the flux tube and its width depend only on $m_0$---that is, on the coefficient of the extrinsic-curvature term of the effective string action.     

It is interesting to compare our findings with the results that have been obtained in non-Abelian gauge theories. It should be noted that, in those theories, $m_0/\sqrt{\sigma}$ is essentially constant (up to small, finite-lattice-spacing corrections): thus the question, whether the flux tube width is a function of $\sigma$ (which appears as the coefficient of the Nambu-Got{\={o}} term in the effective string action) or of $m_0$, is not a well-defined one. By contrast, in compact $\U(1)$ lattice gauge theory in 3D, the issue can be studied in a meaningful way, since in this case 
the $m_0/\sqrt{\sigma}$ ratio is strongly dependent on the coupling:
\begin{equation}
\label{ratio_m0_1}
\frac{m_0}{\sqrt{\sigma}} \simeq \frac{2\pi c_0}{\sqrt{c_\sigma}} (2\pi\beta)^{3/4} e^{-\frac{\pi^2}{2} v(0)\beta},
\end{equation}
so it becomes possible to disentangle the dependence on one quantity from the one on the other.
A natural question that arises, is then if it is nevertheless possible to find evidence of Abrikosov-vortex-like behavior also in $\SU(N)$ gauge theories---and, thus, evidence of non-negligible extrinsic-curvature effects in the string model describing their low-energy dynamics. As $m_0/\sqrt{\sigma}$ is approximately constant, this question cannot be answered by looking at the dependence of the flux-tube width on the interquark distance, but in principle one could address the issue by looking at the flux-tube profile. As figure~\ref{fig:logElvsz-2.0-16a} shows, in our theory the shape of the flux tube at large distance from the interquark axis is unambiguously described by a simple exponential, as predicted by the equations for a dual superconductor in three spacetime dimensions. Unfortunately, the existing examples of this type of analysis for non-Abelian models (for instance, those reported in ref.~\cite{Gliozzi:2010zv} for $\SU(2)$ Yang-Mills theory in 3D, or in refs.~\cite{Cea:2014uja, Cardoso:2013lla} for the $\SU(3)$ theory in 4D) are less conclusive, and appear to reveal a mixture of Gau{\ss}ian and exponential contributions. To get a better picture, it would probably be helpful to address this issue not only numerically but also analytically (for example, by an explicit computation of the flux-tube profile in an effective-string model including an extrinsic-curvature term).

To summarize, in this work we used the dual formulation of compact $\U(1)$ gauge theory in 3D to study the transverse shape of the flux tube on the symmetry line between the static sources. The dual formulation of the theory allowed us to reach high numerical precision, without any error-reduction method.

Fitting our large-transverse-separation results for the flux-tube profile to the prediction of the dual-superconductor model, we found that the London penetration length is compatible with the inverse of the lightest glueball mass $m_0$, and that this quantity is independent of the distance between the static sources, as expected. This prediction, however, can only fit the tails of flux-tube profile.

The squared width $w^2$ of the confining flux tube was evaluated directly from the data obtained in our Monte~Carlo simulations, for a wide range of physical separations $R$ between the color charges. We found that, although $w^2$ grows logarithmically with $R$, as predicted by effective string theory, its amplitude is not proportional to $1/\sigma$ (the inverse of the string tension), but rather to $1/m_0^2$. For non-Abelian Yang-Mills theories in four spacetime dimensions, the ratio of these two dimensionful scales is nearly independent of the lattice spacing (at least at values of the lattice spacing of physical interest, i.e. sufficiently far from the strong-coupling regime), but this is not the case here. In fact, as we already remarked earlier, part of our interest in this lattice model stems precisely from the fact that it allows one to disentangle the dependence of different physical effects on these two scales.

From these results one can conclude that in this theory:
\begin{itemize}
\item the characteristic scale of the flux tube is $1/m_0$ (rather than $1/\sqrt{\sigma}$, as one would expect);
\item at large enough transverse separation, the profile of the flux tube can be successfully modelled by an exponential decrease, in agreement with the expectations from the dual-superconductor model in 3D.
\end{itemize}

The investigation of the structure and dynamics of flux tubes in this toy model of confinement, that we carried out in the present work, could provide guidance toward a better understanding of analogous problems also in non-Abelian theories. 

\acknowledgments

The numerical calculations were carried out on INFN Pisa GRID Data Center machines. We thank Leonardo~Cosmai, Francesca~Cuteri and Michele~Pepe for helpful comments, discussions and correspondence.

\bibliography{paper}

\end{document}